\documentclass[pre,aps,showpacs]{revtex4}

\usepackage{amsmath,amssymb}
\usepackage{graphicx}
\usepackage{psfrag}
\usepackage{graphicx}
\usepackage{epsf}

\newcommand{\Dcrossij}{D^{\times}_{ij}}

\newcommand{\DcrossNullij}{{\Dcrossij}^{(0)}}

\begin{document}

\title{Symmetries and novel universal properties of turbulent hydrodynamics
 in a symmetric binary fluid mixture}

\author{Abhik Basu}

\affiliation{Max-Planck-Institut f\"{u}r Physik Komplexer Systeme,
N\"othnitzer Str. 38, D-01187
Dresden, Germany \cite{byemail}, 
Abteilung Theorie, Hahn-Meitner-Institut, Glienicker Strasse
 100, D-14109 Berlin, Germany, and Poornaprajna Institute of Scientific
Research, Bangalore, India.}

\date{\today}

\begin{abstract}
We elucidate the universal properties of the nonequilibrium steady states
(NESS) in a driven symmetric 
binary fluid mixture, an example of active advection, in its miscible phase. 
We use the symmetries of the equations of motion to
establish the appropriate form of the structure functions which characterise
the statistical properties of the NESS of a driven symmetric binary fluid
mixture.
We elucidate the universal properties described by the scaling
exponents and the amplitude ratios. Our results suggest that these
exponents and amplitude ratios
{\em vary continuously} with the degree of crosscorrelations
between the velocity and the gradient of the concentration fields. 
Furthermore, we demonstrate, in agreement with Celani {\em et al}, [{\em Phys.
Rev. Lett.}, {\bf 89}, 234502 (2002)],
that the conventional structure functions as used in passive scalar turbulence
studies exhibit only simple scaling in the problem of symmetric binary fluid
mixture {\em even} in the weak concentration limit. We also discuss possible
experimental verifications of our results.

\end{abstract}

\maketitle

Critical dynamics of equilibrium systems show universal scaling
properties for thermodynamic functions and correlation functions
characterised by universal scaling exponents that depend on the
spatial dimension $d$ and the symmetry of the order parameter
(e.g., Ising, XY etc.) and conservation laws \cite{fisherrev,hal},
but not on the parameters that specify the details of the
Hamiltonian. A different situation arises in driven, dissipative,
nonequilibrium systems with nonequilibrium statistical steady
states (NESS). In this letter we demonstrate that in the NESS of externally
driven turbulent homogeneous and isotropic symmetric binary fluid mixtures
(50-50 concentration of the two components)
universal properties are characterised by certain structure functions, determined by the symmetries of the equations of motion, and which are different from those used in pasive scalar turbulence studies. We find that the universal properties vary continuously with the degree of
crosscorrelations between the velocity and the gradient of the
concentration fields, 
a situation which has no analogue in equilibrium.

The turbulent hydrodynamics of an incompressible symmetric binary
fluid is described by the coupled evolution of the velocity field
$\bf u({\bf x},t)$ and the  concentration field gradient
$\nabla\psi({\bf x},t)$ with $\psi ({\bf x},t)
=[\rho_A({\bf x},t)-\rho_B({\bf x},t)]/\rho_0$ where $\rho_0$ is
the mean density, and $\rho_A$ and $\rho_B$ are
the densities of the two components. In a {\em symmetric} binary
fluid $\langle \psi({\bf x},t)\rangle =0$. Experiments performed on such
systems have been typically concerned with the measurements of
effective transport coefficients \cite{exp}. In a symmetric binary
mixture, $\psi$ is not simply advected
passively by the $\bf u$-field, but is {\em active}, i.e., concentration
gradient $\nabla \psi$ reacts back on $\bf u$ to produce flows. In ordinary
fluid turbulence the NESS is characterised by the
multiscaling exponents of the structure functions of the velocity
fields (see below)\cite{frisch,rahulrev}. Compared to the studies
of ordinary fluid turbulence the study of turbulent binary
mixtures is in its infancy \cite{exp}.

In this letter we develop a one-loop, self-consistent theory for a
randomly forced symmetric binary mixture system to elucidate the universal 
statistical properties
of its NESS in the miscible region.
We identify the symmetries
of the Eqs. of motion which
allow us to construct the structure functions (different from the ones
discussed in the literature for passive scalar turbulence studies 
- see below) which exhibits multiscaling in the
NESS in the inertial range..
We calculate the interrelations between  certain
dimensionless numbers, 
e.g., the amplitude-ratio of the equal time correlators
of $\bf u$ and $\nabla \psi$ fields and the renormalized
Prandtl number (see below)
in the NESS in $d$-dimensions. Finally, we elucidate the nature of
multiscaling by calculating the multiscaling exponents
for the even order structure functions of the $\bf \nabla \psi$ field in
the limit of small $\psi$  in two different models. In one case we find
the multiscaling exponents to vary with the degree of crosscorrelations between
the $\bf u$ and $\nabla \psi$ fields. We also consider the scaling of the
conventional structure functions and show, in agreement with
Ref.\cite{activesim}, that they exhibit only simple scaling.

\paragraph{Equations of motion, their symmetries and structure functions:-}
It is well-known \cite{activesim,ruiz,jkb1} that in the dynamics of a symmetric
binary fluid mixture the velocity fields $\bf u$ couple 
with the concentration gradient $\nabla\psi$ and not with $\psi$
itself. Thus it is useful to write the coupled evolution equations in terms of
$\bf u$ and $\nabla \psi$. We are interested in the isopropic and homogeneous
situation, i.e., we have no mean concentration gradient, $\langle
\nabla \psi \rangle=0$. Furthermore, for calculational convenience (see below),
we write
$\nabla\psi ={\bf b}({\bf x},t)+{\bf B_o}/{\alpha_o}$,
where $\bf b$ is a space-time dependent vector, 
$\bf B_o$ is a constant vector and $\alpha_o$ is a numerical constant. 
Note that since we
impose $\langle\nabla\psi\rangle=0$ we have $\langle {\bf b}\rangle =
-{\bf B_o}/\alpha_o$. Such
a parametrization of $\nabla \psi$ yields the
Eqs. of motion \cite{activesim,ruiz,jkb1} 
\begin{equation}
\frac{\partial {\bf u}}{\partial t}+\lambda_1({\bf u}\cdot
\nabla){\bf u}= -\frac{\nabla p}{\rho_0}-{\bf B_o}\cdot \nabla
{\bf b}-\alpha_o{\bf b}\nabla\cdot {\bf b} +\nu\nabla^2 {\bf u}
+{\bf f}, \label{navier1}
\end{equation}

\begin{equation}
{\rm and}\,\,\frac{\partial {\bf b}}{\partial t}+\frac{\lambda_2}{\alpha_o}\nabla ({\bf
u\cdot B_o})+\lambda_2{\nabla (\bf u\cdot b)}=\eta\nabla^2 {\bf
b}+{\bf g}. \label{advec2}
\end{equation}
We also consider, for simplicity, the velocity field $\bf u$ to be
incompressible enforced by the condition $\nabla\cdot \bf u=0$.
The parameters $\lambda_1,\,\lambda_2$ and $\alpha_o$ are coupling constants.
Coupling constants $\lambda_1,\,\lambda_2$ are unity and
kept for formal bookkeeping purposes. 
A choice of $\alpha_o$
depends on the unit of $\bf \nabla\psi$ and 
is given by $\alpha_o=\frac{\zeta^2}{m\chi}$
\cite{ruiz} where $\zeta$ is the thermal correlation length, $\chi$ is the
susceptibility per particle, and $m$ is the particle mass. This, within the
mean field theory, yields $\alpha_o$ to be approximately temperature
independent near the critical point $T_c$\cite{ruiz}. On the other hand, 
away from $T_c$, similar considerations yield $\alpha_o\sim T$. Therefore, at
any finite temperature $\alpha_o\sim O(1)$ and, as a result, a symmetric binary
mixture exhibits active behaviour. Functions $\bf f$ and $\bf g$ are
stochastic forces required to maintain a statistical steady state.
Parameters $\nu_o,\,\eta_o$ are the bare viscosity and diffusivity 
respectively. Note that in 
a problem of true passive advection $\alpha_o$ is identically zero. At this
point we would like to clarify that by writing $\nabla\psi$ as the sum of ${\bf
b +B_o}/{\alpha_o}$ 
we do not introduce any anisotropy in the problem as the latter is
decided by $\langle \nabla \psi\rangle =\langle {\bf b}\rangle +{\bf B_o}/{\alpha_o}$ which is
kept zero in our system. Furthermore, stochastic forces $\bf f$ and $\bf g$
have isotropic variances [see Eqs.(\ref{noisecorr})] 
and hence do not violate the isotropicity of the system.

The Eqs. of motion (\ref{navier1}) and (\ref{advec2}) are
invariant under
\begin{itemize}
\item TI: The Galilean transformation ${\bf u}({\bf
x},t)\rightarrow {\bf u}({\bf x+u_o}t,t)+{\bf u_o},
\frac{\partial}{\partial t} \rightarrow \frac{\partial}{\partial
t}- {\bf u_o\cdot \nabla}$ and $\bf b\rightarrow b$ with
$\lambda_1=\lambda_2=1$ \cite{ruiz,jkb1}.

\item TII: The transformation $ {\bf B_o}\rightarrow {\bf
B_o}+\alpha_o\delta,\,{\bf b}({\bf x},t)\rightarrow {\bf b}({\bf
x},t)-\delta,\, {\bf u\rightarrow u}$. Here the shift $\delta$ is a vector
(see, e.g., Ref.\cite{anton} for an analogous symmetry in a related problem).
\end{itemize}
Transformation TII implies that the renormalised versions of the Eqs. 
(\ref{navier1}) and (\ref{advec2}) must be expressible, as the bare equations
themselves, in terms of the total gradient fields 
($={\bf b+B_o}/{\alpha_o}$) \cite{anton}.
It has been overlooked in related previous works
\cite{ruiz,jkb1}. The existence of TI implies nonrenormalisation
of $\lambda_1,\,\lambda_2$. Since the renormalized versions of Eqs.
(\ref{navier1}) and (\ref{advec2}) are are also invariant under the
transformatio TII, the fourth and the fifth terms in Eq. (\ref{navier1}) must
renormalize in the same way. Similarly, 
the second and the third terms in Eq. (\ref{advec2}) must renormalize in 
the same
way. Hence the corresponding renormalization-$Z$ factors must be equal leading
to $Z_{B_o}Z_b=Z_{\alpha_o}Z_b^2$. It can be argued that in model Eqs.
(\ref{navier1}) and (\ref{advec2}) $Z_{B_o}Z_b=1$ \cite{abpassive} (this
essentially is a freedom of choice for the renormalization 
$Z$-factors in the present model). 
This yields $Z_{\alpha_o}=1$.  Here,
$Z_{\alpha_o},\,Z_{B_o}$ and $Z_b$ are
renormalization factors for $\alpha_o,\,\bf B_o$ and $\bf b$
respectively. Therefore, $\alpha_o$ does not renormalize and hence it 
can be set to unity by
treating all the gradient fields as effective gradient fields
$\sqrt \alpha_o{\bf b}$. In fact, this can be argued in a simple
way: Let us assume that under mode elimination and rescaling
$\alpha_o\rightarrow \beta\alpha_o$ where $\beta$,
a scale factor, can be
absorbed by redefining the units of the gradient fields $\bf b$ by
$\sqrt\beta \bf b\rightarrow b$. This leaves Eq.
(\ref{advec2})  and $\alpha_o$ in Eq. (\ref{navier1}) unchanged.
Furthermore, the invariance of the Eqs. motion
(\ref{navier1}) and (\ref{advec2}) under TII ensures that the
composite operators $(b_i)^n$ (Cartesian indices 
$i$ contracted or free) are
also not renormalized \cite{abpassive,adzrev}.
It should be noted that these results hold regardless of the strength
of the feedback of $\bf b$ on $\bf u$. Therefore, even though the
dynamics of the $\bf b$ fields are linear in its {\em passive}
limit, i.e., when the feedback of $\bf b$ on $\bf u$ is ignored,
it must still be consistent with TII and its consequences
\cite{abpassive} for it to 
represent the physical weak scalar limit of
the binary mixture problem. We emphasize that by the 
transformation TII we do not introduce a
mean concentration gradient $\langle \nabla \psi\rangle$ in the system as 
the latter, being given by ${\bf
B_o} +\langle {\bf b}({\bf x},t)\rangle$ is independent of such shifts. In
particular we work with 
$\langle \nabla \psi\rangle$ =0 (i.e., no mean concentration gradient).
This, furthermore, can be achieved in a perturbative calculation by (i)
$\langle {\bf b}({\bf x},t)\rangle = -{\bf B_o}\neq 0$, or, (ii)
$\langle {\bf b}({\bf x},t)\rangle = -{\bf B_o}=0$, both of which are in
agreement with $\langle \nabla \psi\rangle =0$. The invariance under the
transformation TII ensures that physical results are independent of such
choices. In our calculations below we work
with $\langle {\bf b}({\bf x},t)\rangle =
-{\bf B_o}=0$ which is the most convenient one.

In fluid turbulence, the NESS is characterized by the multiscaling
of the structure functions $S_n^u(r)=\langle [|u_i({\bf x
+r})-u_i({\bf x})|\hat r_i]^n\rangle\sim r^{\zeta_n^u}$ for $r$ in
the inertial range (i.e., $\eta_D\ll r\ll L,\,\eta_D$ is the
dissipative cutoff scale, $L$ is the integral scale $\sim$ system
size) which are invariant under the  Galilean transformation.
Analogously, in the
problem of symmetric binary fluid mixtures, the appropriate structure
functions must be invariant under TI and TII. Such functions are
$S_n^a(r)=\langle [|a_i({\bf x +r})-a_i({\bf x})|\hat
r_i]^n\rangle,\,a=u,b$. 
The scalar field structure functions as used in passive scalar
turbulence studies \cite{passive}, 
$S^{\psi}_n=\langle [\psi({\bf x +r})-\psi({\bf
x})]^n\rangle$ are not invariant under TII: ${\bf b\rightarrow b} +
constant$.
and exhibit only simple scaling in the symmetric binary mixture problem (
and hence not suitable for the studies of the passive limit of the
active problem), as
we show below and has been observed in Ref.
\cite{activesim}. In this communication we calculate the
multiscaling exponents of the even order structure functions
$\mathcal{S}^b_{2n}(r)=\langle [b_i({\bf x+r})-b_i({\bf
x})]^{2n}\rangle$ by using one-loop dynamic renormalisation group
(RG) in a field theoretic framework
in the weak scalar limit of the
symmetric binary mixture problem.

\paragraph{Correlation functions, scaling exponents and
dimensionless numbers:-}

Since $\bf u$ and $\bf b$ are polar and axial vectors (Note: a mass density
is a pseudo scalar in $3d$ and hence $\bf b=\nabla \psi$ is an axial
vector)
respectively, the crosscorrelation tensor $C_{ij}^{\times}=\langle
u_i ({\bf k},t)b_j ({\bf -k},0)\rangle$ is imaginary and odd in the
wavevector $\bf k$. The propagators 
$G_a,\,a=u,b$ of $\bf u$ and $\bf b$ 
as functions of $\bf k$ and frequency
$\omega$ are defined as 
\begin{equation}
u_i({\bf k},\omega)\equiv 
G_u(k,\omega)f_i({\bf k},\omega),b_i({\bf k},\omega)\equiv 
G_b(k,\omega)
g_i({\bf k},\omega),
\end{equation}
 and the correlators 
\begin{equation}
C^u_{ij}(k,\omega)=\langle u_i({\bf k},\omega)u_j({\bf
-k},-\omega)\rangle,\,C^b_{ij}(k,\omega)=
\langle b_i({\bf k},\omega)b_j({\bf
-k},-\omega)\rangle. 
\end{equation}
In the scaling limit, in terms of the dynamic
exponent $z$ and the roughness exponents $\chi_u,\,\chi_b$ of the
fields $\bf u$ and $\bf b$, 
$G_u^{-1}=-i\omega-\Sigma_u(k,\omega),\,
G_b^{-1}=-i\omega-\Sigma_b(k,\omega)$ where self-energies
$\Sigma_u=k^{-z}m_u(\omega/k^z),\,\Sigma_b=k^{-z}m_b(\omega/k^z)$
and the correlators
$C_{ij}^u=P_{ij}({\bf k})k^{-d-2\chi_u-z}m_1(\omega/k^z),\,C_{ij}^b=
Q_{ij}({\bf k})k^{-d-2\chi_b-z}m_2(\omega/k^z)$
with $P_{ij}({\bf k})=\delta_{ij}-\frac{k_ik_j}{k^2}$ and 
$Q_{ij}=\frac{k_ik_j}{k^2}$ being the transverse and longitudinal 
projection operators respectively. The Ward
identities arising from TI and TII yield $\chi_u=\chi_b=\chi$ and
$\chi+z=1$. 

We assume zero-mean Gaussian distribution for $\bf f$ and $\bf g$
with variances
\begin{eqnarray}
  \langle f_i ({\bf k},t) f_j ({-{\bf k},0}) \rangle
          &=&  2 D^f_{ij}({\bf k}) \delta (t), \nonumber \\
 \langle g_i ({\bf k},t) g_j ({-{\bf k},0}) \rangle
          &=&  2 D^g_{ij}({\bf k}) \delta (t), \nonumber \\
    \langle f_i ({\bf k},t) g_j ({-{\bf k},0}) \rangle
          &=&  2 i {\DcrossNullij} ({\bf k}) \delta (t),
\label{noisecorr}
\end{eqnarray}
where $D^{f,g}_{ij}({\bf k})=D^{f,g}_0H_{ij}({\bf
k})k^{4-\epsilon -d},\,\epsilon>0,\,H_{ij}=P_{ij},Q_{ij}$ for $f,g$
respectively. 
Specifically, $\epsilon=4$ reproduces
Kolmogorov-type (hereafter K41) 
scaling for the kinetic energy spectrum [$k^{d-1} \langle {\bf u}({\bf
k},t)\cdot {\bf u}({\bf -k},t)\rangle$] and the gradient energy spectrum
[$k^{d-1}
\langle {\bf b}({\bf k},t)\cdot {\bf b}({\bf -k},t)\rangle$]
\cite{jkb1}. Allowing ${\DcrossNullij} ({\bf k})$,
(odd in $\bf k$), to break rotational invariance or symmetry with
respect to an interchange of the cartesian indices $i$ and $j$ we
write ${\DcrossNullij} ({\bf k})={\DcrossNullij}_s ({\bf
k})+{\DcrossNullij}_a ({\bf k})$, with ${\DcrossNullij}_s$ and
${\DcrossNullij}_a$ being the symmetric and the antisymmetric
parts of ${\DcrossNullij}$. We choose ${\DcrossNullij}_s ({\bf
k})= D_s^o P_{ij}({\bf k})k^{4-\epsilon -d},\, {\DcrossNullij}_a
({\bf k})=D_a^o\epsilon_{ijp}k_pk^{-\epsilon},\,\epsilon_{ijp}$ is the
totally antisymmetric tensor in $3d$ \cite{foot1}. 
A one-loop self-consistent mode coupling (SCMC)
theory is conveniently written in terms of the self-energies and
the correlators involved.

We start with the zero frequency ($\omega=0$) forms for the self-energies
\begin{equation}
\Sigma_u(k,\omega=0)=\nu
k^z,\,\Sigma_b(k,\omega=0)=\eta k^z,
\end{equation}
 and the correlators
\begin{eqnarray}
C_{ij}^u(k,\omega=0)&=&\frac{2D_1}{\nu^2}P_{ij}k^{-2\chi-d-z},\nonumber \\
C_{ij}^b(k,\omega=0)&=&\frac{2D_2}{\eta^2}Q_{ij}k^{-2\chi-d-z},\nonumber \\
C_{ij}^s(k,\omega=0)&=&\frac{2iD_s({\bf
k})}{\nu\eta}P_{ij}k^{-2\chi-d-z},\,D_s({\bf k})=-D_s({\bf -k})
\end{eqnarray}
in $d$ dimensions and
\begin{equation}
C_{ij}^a=\frac{2iD_a}{\nu\eta}\epsilon_{ijp}k_p k^{-2\chi-z-4}
\end{equation}
in $3d$. Subscripts $s$ and $a$ refer to the symmetric and
antisymmetric parts of $C_{ij}^{\times}$.
The lack of vertex
renormalizations in the present problem in the zero wavevector
limit implies that the SCMC yields exact relations between $\chi$ and $z$,
viz, $\chi+z=1$ (see, e.g., \cite{abepl}). By matching the
self-energies and the correlation functions calculated at the
one-loop order at $\omega=0$ with
their zero frequency expressions above (we take
$\lambda_1=\lambda_2=\alpha_o=1$ without any loss of generality)
we obtain 
\begin{equation}
z=2-\epsilon/3,\,\chi=-1+\epsilon/3.
\end{equation}
Furthermore, we
obtain self-consistent relations 
between the two dimensionless numbers, the  amplitude-ratio $\Gamma=D_2/D_1$ and
the renormalized Prandtl number
$P=\eta/\nu$. We consider the cases of symmetric and
antisymmetric crosscorrelations separately \cite{abepl}. When
$D_a=0$, i.e., in the absence of antisymmetric crosscorrelations,
we find
\begin{equation}
P^{-1}=\frac{\frac{d-1}{2(d+2)}+\frac{\Gamma}{2P^2(d+2)}}{\frac{d-1}{d(1+P)}
-\frac{\Gamma(d-1)}{Pd(1+P)}}, \label{eqpm}
\end{equation}
\begin{equation}
\Gamma=\frac{\frac{d^2-2}{d(d+2)}+\frac{\Gamma^2}{P^3d(d+2)}-\frac{4(d^2-2)N_s^2}
{P(1+P)}}{\frac{2\Gamma^{-1}
(d-1)}{P(1+P)d}+\frac{2N_s^2(d-1)}{P(1+P)\Gamma^2d}},
\label{eqgam1}
\end{equation}
where $N_s=\left[\frac{\langle u_i({\bf k},t)b_i({\bf
-k},t)\rangle}{\langle u_i({\bf k},t)u_i({\bf
-k},t)\rangle}\right]^2$, parametrising  the strength of symmetric
crosscorrelations in the system.

Similarly, for finite $D_a$, i.e., finite antisymmetric
crosscorrelations and with $D_s=0$, Eq. (\ref{eqpm}) remains
unchanged (meaningful in $3d$ only) but Eq. (\ref{eqgam1})
changes to
\begin{equation}
\Gamma=\frac{\frac{7}{15}+\frac{\Gamma^2}{45P^3}+\frac{4N_a^2}{3P(1+P)}}
{\frac{4\Gamma^{-1}}{3P(1+P)}-\frac{4N_a^2}{P(1+P)}},
\label{eqgam2}
\end{equation}
with $N_a=\left[\frac{\langle u_i({\bf k},t)b_j({\bf
-k},t)\rangle_a \langle u_i({\bf k},t)b_j({\bf -k},t)\rangle_a }
{\langle u_i({\bf k},t)u_i({\bf -k},t)\rangle^2}\right]$
parametrising the strength of antisymmetric crosscorrelations in
the system (subscript $a$ refers to the antisymmetric part). 
Note that both $N_s$ and $N_a$ are dimensionless
numbers since $\chi_u=\chi_b$. Our results imply that in
the NESS, the values of $\Gamma$
and $P$ (i) are related to each other and (ii)they depend on
the dimensionless numbers $N_s,\,N_a$, i.e., the degree of the
crosscorrelations between $\bf u$ and $\bf b$.
It is significant that similar variations of the
dimensionless numbers and their relations with the breakdown of the invariances
under rotation and the interchange of the cartesian indices, are observed 
also in MHD
turbulence \cite{abepl}. 
This strongly suggests that such properties, having no equilibrium
analogue, are rather generic properties of a class of nonequilibrium systems.
We close this Section by noting that the parameters $N_s$ and $N_a$ being the
dimensionless ratios of aprropriate correlators in the NESS are
measurable in experiments and numerical simulations, and not the same as the
amplitudes of noise crosscorrelations (\ref{noisecorr}) which are introduced 
in our framework as
convenient means to calculate different correlation functions.

\paragraph{Multiscaling exponents $\zeta_{2n}^b$:-}
A full characterisation of the NESS for a turbulent binary mixture
requires the calculations of $S_n^a(r),\,a=u,b$ (see above) which
scale as $\sim r^{\zeta^a_n}$ for $r$ in the inertial range. There
is no controlled perturbative calculation which permits to do this to
this date. Complications of the full nonlinear problem initiated
studies of multiscaling for the gradient field in the passive
advection limit where the backreactions of the gradient field on
the dynamics of the velocity field are neglected. There is a large
body of literature on this problem, known as the {\em passive
scalar turbulence}, see, e.g., \cite{passive}. However we
differ from others in ensuring that our calculations obey the symmetry
TII which was not considered in the existing works. 

K41-type arguments for fluid turbulence \cite{k41}
 can be easily extended to a binary mixture system
by demanding that the scaling in the inertial range is determined
by the local length scale $r$ and the rate of mean dissipation of
energy (kinetic or gradient). This yields, $\zeta_n^a=n/3$ as in
fluid turbulence. However, departures from the $K41$ scaling (or
multiscaling) in fluid turbulence suggest the presence of similar
features in the binary mixture problem. Below we analyse the
scaling properties of the even order structure functions
${\mathcal S}_{2n}^b(r)=\langle [b_i({\bf x+r})-b_i({\bf
x})]^{2n}\rangle$ in the passive limit of the full nonlinear problem.
Further, the velocity field $\bf u$ is assumed to obey a
Gaussian distribution with zero mean and a variance \cite{passive}
\begin{equation}
\langle u_i({\bf k},t)u_j({-\bf k},0)\rangle= {\frac{D_oP_{ij}(k)}
{(k^2+m^2)^{ d/2+\varepsilon /2}}}\delta (t), \label{kraichvari}
\end{equation}
where $0<\varepsilon <2$. The field $\bf b$ is then governed by
the advective equation
\begin{equation}
\frac{\partial {\bf b}}{\partial t}+\lambda_2{\nabla (\bf u\cdot
b)}=\eta_o\nabla^2 {\bf b}+{\bf T}. 
\label{advec3}
\end{equation}
 Here $\bf T$ is a short range force with correlations
$\langle T_i({\bf x+r},t)T_j({\bf x})\rangle=Q_{ij}A(mr)\delta
(t)$
with $m\sim 1/L,\,L$ being the system size and $A(r/L)\rightarrow
const.$ for $L\rightarrow \infty$ \cite{passive}. Note that following the
formulations of Ref.\cite{passive} the longrange 
stochastic forcing function $\bf g$ in Eq. (\ref{advec2}) is replaced by the
short ranged force $\bf T$ in Eq. (\ref{advec3}). We continue to impose the
consequences arising from TII. 
Thus it can be called the {\em modified} Kraichnan
model, modified to study the weak concentration gradient
limit of the {\em active} problem.

We employ a one-loop dynamic renormalisation group analysis with a
minimal subtraction scheme in our treatment \cite{adzrev}. Following standard
procedures \cite{abpassive,adzrev,amit} we obtain the critical
dimension of the fields $\bf b$ to be given by $\Delta_b=-1+\varepsilon/2$
\cite{abpassive},
where $\varepsilon$ is defined through the relation (\ref{kraichvari}).
Hence, from the definition of critical dimensions we find
that the structure functions ${\mathcal S}_{2n}^b$ scale as
${\mathcal S}_{2n}^b(r)=r^{n(2-\varepsilon)}f_n(mr),$
with yet undetermined scaling functions $f_n(mr)$. Anomalous
scaling or {\em multiscaling} would arise if the scaling functions
were to behave in the asymptotic limit $r/L \rightarrow 0$ as
$(r/L)^{\delta_n}$ (for $\delta_n=0$ simple scaling would
prevail). Ignoring such possibilities for the time being we find
that $\zeta_{2n}^b(K41)=n(2-\varepsilon)$ which is the analog of
K41 scaling in fluid turbulence in this problem \cite{abpassive}.

We now turn to the study of the 
asymptotic forms of the scaling functions for $mr\ll 1$, i.e.,
in the inertial range $\eta_D\ll r\ll L$ by using
renormalisation of the most relevant composite operators in Wilson
Operator Product Expansions (OPE)\cite{amit} of the scaling
functions. This method has already been used in critical
phenomena \cite{amit}, fluid turbulence \cite{adzrev} and passive
scalar turbulence \cite{passive}. The anomalous dimensions
$\gamma_n$ of the most relevant composite operators for $O_n=[b_i({\bf
x+r})-b_i({\bf x})]^{2n}$
are calculated in a one-loop $\varepsilon$-expansion. This
then provides the deviation from simple scaling: ${\mathcal
S}_{2n}^b(r)\sim r^{n(2-\varepsilon)+\tau_n}$ where $\tau_n$ is
the anomalous dimension of $O_n$ coming from the composite
operators. In the OPE of $O_n$
only renormalised operators which are invariant under TI and TII
(which keep $O_n$ invariant) can appear. Moreover, in the
passive limit the linearity of Eq.
(\ref{advec2}) ensures that the number of $\bf b$ fields appearing
in the OPE of $O_n$ cannot exceed $2n$ \cite{abpassive}. For $O_2$ the leading
operator is $[\partial_i b_j({\bf x})\partial_i b_j ({\bf x})]$
with an anomalous dimension $\gamma_2=\varepsilon$, leading to
$\zeta_2^g=2>\zeta_2^b(K41)$. For $n>1$ the most relevant operator
in the OPE of $[\partial_i b_j({\bf x})]^p[\partial_i b_j({\bf
x'})]^q,\,p+q\leq 2n$, which determines the anomalous dimension of
$O_n$, is $(\partial_i b_j\partial_i b_j)^n$. This yields \cite{abpassive}
\begin{equation}
\zeta_{2n}^b=n(2-\varepsilon)-\frac{2n^2+nd}{d+2}\varepsilon<\zeta_{2n}^b(K41).
\end{equation}
Note that our results satisfy  key characteristics of multiscaling
as observed in fluid turbulence studies \cite{frisch,rahulrev}:
$\zeta_{2n}^b$ are a convex function of $n$ and
$\zeta_2^b>\zeta_2^b(K41),\,\zeta_{2n}^b<
\zeta_{2n}^b(K41),\,n>1$. The structure functions for the velocity
fields $S_n^u$ do not multiscale in the present approximation due
to the assumed Gaussian distribution of the velocity fields.

In the above we assumed the velocity fields to be Gaussian
distributed. In contrast, real velocity fields follow Eq.
(\ref{navier1}) which we now consider. 
We still ignore the backreaction of $\bf b$ on the
dynamics of $\bf u$; nevertheless, our calculations are consistent
with the requirements imposed by TII. We are able to obtain
additional results concerning the multiscaling properties
of the structure functions despite the limitations
\cite{abpassive,adzrev} of a one-loop $\epsilon$-expansion method
in this problem where $\epsilon$ is now defined through the noise
correlations (\ref{noisecorr}). We follow the calculational scheme
outlined above and elucidate the multiscaling properties of
${\mathcal S}_{2n}^b$ in a one-loop $\epsilon$-expansion. In the
present problem the critical dimension of the field $\bf b$ is
$(2-\epsilon/3)/2$, yielding for the structure functions
\begin{equation}
{\mathcal S}_{2n}^b \sim r^{n(2-\epsilon/3)}f_n(mr).
\end{equation}
Here 
$\epsilon=4$ is the physically interesting case.
In order to find out any deviation from
this simple scaling we examine the asymptotic form of $f_n(mr)$ in
the procedure outlined above which yields the following results: Writing
${\mathcal S}^b_{2n}(r)\sim r^{n(2-\epsilon/3)+\tau_n}$ we obtain
\cite{abpassive}
\begin{equation}
\tau_1= \frac{\epsilon}{3P(1+P)}[1+\frac{4P^2}{1+P^2}]
\end{equation}
and for $n>1$,
\begin{equation}
\tau_n=-\frac{n(n-1)
\epsilon}{3P(1+P)d}[\frac{d-1}{d+2}+\frac{140P^2(4-d) (d-1)}
{(d+2)(d+4)(d+6)}<0
\end{equation} 
making $\zeta_{2n}^b<\zeta_{2n}^b(K41)$, where $P$ is the
{\em renormalised Prandtl number}.
Thus, again, exponents $\zeta_{2n}^b$ follow the same key
characteristics of multiscaling as observed in fluid turbulence.
The most striking aspect of our results for $\zeta_{2n}^b$ is their
dependence on $P$.
According to our results obtained above 
$P$ is not a fixed number in the NESS of a binary fluid mixture, but rather
parametrised by the degree of crosscorrelations between $\bf u$
and $\bf b$. Even though our results for $\zeta_{2n}^b$ are
obtained in the passive limit 
their dependences on $P$ are likely to remain in the full
nonlinear problem, and
thereby open up the intriguing possibility of variable
multiscaling parametrised by the degree of crosscorrelations
between $\bf u$ and $\bf b$.

Finally we consider the scaling of the structure functions
$S_{2n}^{\psi}=\langle [\psi({\bf x+r})-\psi ({\bf
x})]^{2n}\rangle$, commonly discussed in the passive scalar
literature \cite{passive}. We have $S_{2n}^{\psi}(r)\sim
r^{n(4-\epsilon)}\phi_n(mr)$ in the case where the $\bf u$ fields
are assumed to have a Gaussian distribution with zero mean and a
variance given by (\ref{kraichvari}).  
The asymptotic limit of the scaling functions $\phi_n(mr)$ 
will determine the multiscaling of $S_{2n}^{\psi}(r)$, which are not
invariant underi the transformation TII. Hence in the OPE
of $O_{\psi}^n=[\psi({\bf x+r})-\psi ({\bf x})]^{2n}$ operators violating TII
may appear \cite{abpassive}. 
By virtue of TII operators
$(b_i)^{2n}$ (most dominant in the OPE of $O_{\psi}$)
do not renormalize and as a result the anomalous
dimension of $O_{\psi}^n$ is zero \cite{abpassive}.
Therefore, $S_{2n}^{\psi}(r)\sim r^{n(4-\epsilon)}$ exhibiting
only simple scaling as observed in Ref.\cite{activesim} for $2d$.
This conclusion remains unaffected even when
$\bf u$ is to be given by the solutions of the Navier-Stokes equation.
Therefore our results provide a symmetry motivated understanding
for those in Ref.\cite{activesim}. It should be noted that in the problem of
true passive advection, e.g., the advection of a temperature field by an
incompressible field, $\alpha_o=0$ and as a result, the condition imposed by
TII does not exist any more. Therefore, the operators $(b_i)^n$ renormalise and
contribute to the multiscaling of $S_{2n}^{\psi}(r)$ as has been found in
Refs.\cite{passive}.

In summary, we have discussed the interplay between the
symmetries of the model equations of a binary mixtures and its
scaling and multiscaling properties in the NESS. We have highlighted the
dependence of the statistical properties of the NESS on  the
degree of crosscorrelations between the velocity and gradient
fields. Our results open up the intriguing possibility of
continuously varying multiscaling universality class for such a
system. Our results on multiscaling suggest the {\em singular nature} of the
weak scalar limit of the full nonlinear problem and connects the presence of a
symmetry to such properties. 
Our results display the key
aspects of multiscaling as observed in fluid turbulence. 
Although the noise crosscorrelations (\ref{noisecorr}) 
implies that the velocity field $\bf u$ and the concentration 
gradient field $\bf b$ have symmetric and antisymmetric crosscorrelations
$\bf u$ and the concentration field $\psi$ have {\em no} crosscorrelations.
Crosscorrelations between $\bf u$ and $\psi$ can arise only if fields $\bf u$ 
are compressible. 
It should be noted that while
comparing with real data one needs to impose the irrotational condition with
respect to the second index in the third of the Eqs. (\ref{noisecorr}). This
arises due the fact the vector $\bf b$ is purely solenoidal (${\bf b}=\nabla
\psi$). This yields a relation between $N_s$ and $N_a$; however, one of them 
still remains free and allows for variable scaling properties in the steady
state. Such properties are essentially nonequilibrium since in equilibrium
noise crosscorrelations vanish identically. Our results may be verified in
turbulent experiments with symmetric 
binary mixtures in their miscible phase. In particular we suggest experiments
active-grid generated turbulence experiments
\cite{war} with symmetric binary fluid in its miscible phase. Such
active-grid experimental set-ups, which, to a good extent, are 
practical realizations of
stochastically driven turbulence models, are fitted with grids with agitator 
wings which rotate with frequencies and for durations which are randomly chosen
with given distributions. Note that in the present case, crosscorrelation
functions of the velocity and the
concentration gradient fields are odd under parity reversal which are absent in
equilibrium and can be generated only by appropriate 
external drives. 
In experimental set-ups such
spatial randomness can be incorporated by adding several layers of active
grids, each layer having random distributions of agitator wings. The effective
stochastic forcings, generated by
allowing agitator wings belonging to different levels of grids to have
different rotation speeds and time durations  with the spatial distributions of
agitator wings breaking inversion symmetry will generate non-zero
crosscorrelation functions of the velocity and the concentration gradient field
at the lowest order. In the resulting nonequilibrium steady state, the
concentration fluctuations can be studied with light-scattering techniques
\cite{light}.
In numerical experiments such crosscorrelations can be generated by
extending the methods of A. Sain {\em et al} \cite{rfnse} by controlling the
appropriate Grashof numbers \cite{gras}, constructed out of the various
noise correlators.
In the closing we would like to comment that since the $\alpha_o$ term in Eq.
(\ref{navier1}) is $O(1)$ in symmetric binary mixtures, 
experimental studies will reveal the active nature of the dynamics, and more
specifically, measurements of multiscaling should verify that the structure
functions $S_{2n}^b(r)$ exhibit multiscaling, where as $S_{2n}^{\psi}(r)$
exhibit simple scaling {\em even} in the weak concentration field limit.
Furthermore, in asymmetric binary mixtures the
correlation length $\zeta$ is small and hence, in the relevant scaling regime
over length scales $\gg \zeta$ the $\alpha_o$ term in Eq. (\ref{navier1}) is
very small leading to experimental situations revealing passive behaviour.
From a wider point of view our results illustrate the richness of
nonequilibrium phenomenologies. Recently, analyses on model
nonequlibrium systems \cite{drossel} revealed the existence of
continuous variation of the scaling exponents. Ours is an
illustration of such variations of 
multiscaling universality classes in natural systems. 
Our results suggest further studies 
for similar effects in driven soft-matter and biologically relevant
systems.

The author wishes to thank J. Santos for 
discussions, D. R. Nelson for critical comments and Alexander
von Humboldt Foundation, Germany for partial financial support.


\begin{thebibliography}{99}
\bibitem[*]{byemail}Present address, e-mail: abhik@mpipks-dresden.mpg.de
\bibitem{fisherrev}M. E. Fisher, in {\em Critical Phenomena, Lect.
Notes Phys.}, Springer Verlag, Berlin, (1983).
\bibitem{hal}P. C. Hohenberg {\em et al}, {\em Rev. Mod.
Phys.}, {\bf 49}, 435 (1977).
\bibitem{exp}H. L. Swinney {\em et al.}, {\em Phys. Rev. A},
{\bf 8}, 2586 (1973).
\bibitem{frisch}U. Frisch, {\em Turbulence: The Legacy of A.N. Kolmogorov}, Cambridge University Press,
Cambridge (1995).
\bibitem{rahulrev} S. K. Dhar {\em et al.},
{\em Pramana - J. Phys.}, {\bf 48}, 325 (1997).
\bibitem{activesim} A. Celani {\em et al}, {\em Phys. Rev. Lett.}, {\bf 89},
234502 (2002).
\bibitem{ruiz}R. Ruiz and D. R. Nelson, {\em Phys. Rev. A}, {\bf
23}, 3224 (1981).
\bibitem{jkb1}M. K. Nandy {\em et al.}, {\em J. Phys.
A}, {\bf 31}, 2621 (1998).
\bibitem{anton}See, e.g., N. V. Antonov, {\em Physica D}, {\bf 144}, 370
(2000).
\bibitem{abpassive}For some details, see A. Basu, unpublished.
\bibitem{passive}M. Chertkov {\em et al}, {\em Phys. Rev. E}, {\bf 52}, 4924
(1995); M. Chertkov {\em et al}, {\em Phys. Rev. Lett.}, {\bf 76},
2706 (1996); K. Gawedzki and A. Kupiainen, {\em Phys. Rev. Lett.},
{\bf 75}, 3834 (1995); D. Bernard {\em et al.} {\em Phys. Rev. E},
{\bf 54}, 2564 (1996); L. Ts. Adzhemyan {\em et al.} {\em Phys.
Rev. E}, {\bf 58}, 1823 (1998).
\bibitem{abepl}A. Basu, {\em Europhys. Lett.}, {\bf 65}, 505
(2004).
\bibitem{k41}A. N. Kolmogorov, {\em C. R. Acad. Sci. USSR}, {\bf
30}, 301 (1941).
\bibitem{adzrev}L. Ts. Adzhemyan {\em et al.},{\em The
Field Theoretic Renormalization Group in Fully Developed
Turbulence}, Gordon and Breach, London (1999).
\bibitem{foot1}This form for $3d$ can be generalized to
arbitrary dimensions $d$ in standard ways.
\bibitem{amit}J. Zinn-Justin, {\em Quantum Field Theory and Critical
Phenomena},
Clarendon, Oxford (1989).
\bibitem{war}R. E. G. Poorte and A. Biesheuvel, {\em J. Fluid Mech.}, {\bf
461}, 127 (2002), A. Gylfason and Z. Warhaft, {\em Phys. Fluids}, {\bf 16},
4012 (2004), and references therein.
\bibitem{light}See, e.g., W. I. Goldberg and J. S. Huang, in {\em Fluctuations,
Instabilities and Phase Transitions}, edited by T. Riste (Plenum, New York,
1975).
\bibitem{rfnse}A. Sain {\em et al}, {\em Phys. Rev. Lett.}, {\bf 81}, 4377 
(1998).
\bibitem{drossel}B. Drossel {\em et al.}, cond-mat/0307579; A.
Basu and E. Frey,{\em Phys. Rev. E}, {\bf 69}, 015101(R) (2004).
\bibitem{gras}C. R. Doering J. D. Gibbon, {\em Applied Analysis of the
Navier-Stokes equation}, Cambridge University Press, Cambridge (1995).
\end{thebibliography}
\end{document}